\documentclass[a4paper,10pt,twoside]{cpc-hepnp}

\usepackage{multicol}
\usepackage{graphicx}
\usepackage{booktabs}
\usepackage{amssymb,bm,mathrsfs,bbm,amscd}
\usepackage[tbtags]{amsmath}
\usepackage{lastpage}
\usepackage{epstopdf}
\usepackage{multirow}

\def\degree{${}^{\circ}$}

\begin{document}
%\begin{CJK*}{GBK}{song}

%\fancyhead[c]{\small Chinese Physics C~~~Vol. 37, No. 1 (2013)010201} \fancyfoot[C]{\small 010201-\thepage}

\fancyhead[c]{\small Submitted to Chinese Physics C} \fancyfoot[C]{\small \thepage}

%\footnotetext[0]{Received 14 March 2014}
\footnotetext[0]{*Supported by National Science Foundation of China(11225525, 11390384)}

\title{Measurement of the liquid scintillator nonlinear energy response to electron}

\author{%
      ZHANG Fei-Hong$^{1,2;1)}$\email{zhangfh@ihep.ac.cn}%
\quad YU Bo-Xiang$^{1;2)}$\email{yubx@ihep.ac.cn}%
\quad HU Wei$^{1,2}$
\quad Yang Ma-Sheng$^{1,2}$\\
\quad CAO Guo-Fu$^{1}$
\quad CAO Jun$^{1}$
\quad ZHOU Li$^{1}$
}
\maketitle

\address{%
$^1$ State Key Laboratory of Particle Detection and Electronics, (Institute of High Energy Physics, CAS) Beijing 100049, China\\
$^2$ University of Chinese Academy of Sciences, Beijing 100049, China\\
}

\begin{abstract}
Nonlinearity of the liquid scintillator energy response is a key to measuring the neutrino energy spectrum in reactor neutrino experiments such as Daya Bay and JUNO. We measured in laboratory the nonlinearity of the Linear Alkyl Benzene based liquid scintillator, which is used in Daya Bay and will be used in JUNO, via Compton scattering process. By tagging the scattered gamma from the liquid scintillator sample simultaneously at seven angles, the instability of the system was largely cancelled. The accurately measured nonlinearity will improve the precision of the $\theta_{13}$, $\Delta m^2$, and reactor neutrino spectrum measurements at Daya Bay.

\end{abstract}

\begin{keyword}
liquid scintillator, nonlinearity, Compton scattering
\end{keyword}

\begin{pacs}
14.60.Pq, 12.15.Ff
\end{pacs}

\footnotetext[0]{\hspace*{-3mm}\raisebox{0.3ex}{$\scriptstyle\copyright$}2013
Chinese Physical Society and the Institute of High Energy Physics
of the Chinese Academy of Sciences and the Institute
of Modern Physics of the Chinese Academy of Sciences and IOP Publishing Ltd}%

\begin{multicols}{2}

\section{Introduction}

Reactor neutrino experiments with a large liquid scintillator (LS) detector play important roles in
neutrino studies. KamLAND experiment first observed the reactor antineutrino disappearance~\cite{kamland}. CHOOZ~\cite{chooz} and Palo Verde~\cite{paloverde} experiments found that the third neutrino mixing angle $\theta_{13}$ is much smaller than the other two. Recently Daya Bay~\cite{dyb_discovery,dyb_cpc,dyb_shape}, Double Chooz~\cite{dc}, and RENO~\cite{reno} discovered an unexpectedly large $\theta_{13}$.

The energy spectrum distortion of the reactor neutrinos at different distances from the reactor(s) is a distinct signal of the neutrino oscillation. However, the energy response of the LS is not linear and is particle-dependent. To precisely measure the energy spectrum, the nonlinearity of the LS detector has to be determined.

The reactor neutrino energy in a LS detector is determined via the observed positron energy in the inverse $\beta$-decay reaction, $\overline{\nu}+p\rightarrow {e}^++n$. The positron losses energy via ionization, and finally annihilate with an electron into two 0.511 MeV $\gamma$s. The $\gamma$s then losses energy via Compton scattering. The Compton electrons further deposit energy in LS via ionization. The LS nonlinearity is related to intrinsic scintillator
quenching and Cherenkov light emission. The latter is affected by the complex absorption and reemission of photons in the LS, and is thus difficult to simulate accurately. The ionization energy of the positron and electron can be assumed to have the same nonlinearity. The gamma energy deposit can be treated as the sum of a series of Compton electron. Since the scintillator quenching is energy-dependent, the energy response of the LS will be different for $e^-$, $e^+$, and $\gamma$ of the same energy. Unfortunately it is not easy to calibrate the detector with $e^-$ or $e^+$ of given energy, especially at multiple locations in the detector. In Daya Bay, multiple $\gamma$ sources, $\beta$ spectrum of the cosmogenic isotope $^{12}B$, as well as $\gamma$s from radioactivity backgrounds and neutron capture on nuclei were used to determine the nonlinearity of the detector. However, the LS nonlinearity is entangled with the nonlinearity of the electronics readout~\cite{dyb_shape}. The nonlinearity for the $e^+$ (and thus the neutrino energy spectrum) determined from the $\gamma$s and the $^{12}B$ carries relatively large uncertainties.

In this study, we measure in laboratory the nonlinearity of the Daya Bay LS for electron via Compton scattering process. With this measurement, we can obtain the nonlinearity for $\gamma$ and positron since the electron response is more fundamental and electromagnetic process can be simulated accurately. Combined with the in-situ $\gamma$s calibrations and other data in detector, the accurately measured nonlinearity will improve the precision of the $\theta_{13}$, $\Delta m^2$, and reactor neutrino spectrum measurements at Daya Bay. The future Jiangmen Underground Neutrino Observatory (JUNO) relies on energy spectrum measurement to determine the neutrino mass hierarchy. It will use similar LS as Daya Bay, although not gadolinium-doped. The LS consists of linear alkylbenzene (LAB) as the solvent, 3 g/L 2,5-diphenyloxazole (PPO) as the primary fluorescence material, and 15 mg/L p-bis-(o-methylstyryl)-benzene (bis-MSB) as the wavelength shifter. This study will also help to understand the energy response of the JUNO detector.

\section{Experimental setup}

The measurement was designed to use the Compton scattering of $\gamma$ of known energy to produce mono-energetic electrons in the LS by tagging the scattered $\gamma$ at certain angles. Fig.~\ref{figsetup} shows the experimental setup. We used a 0.3 mCi $^{22}$Na source, which emits $\gamma$ rays of 0.511 MeV and 1.275 MeV. After passing through a lead collimator with a 9 mm hole on it, $\gamma$s scattered in the LS, which was held in a cylindrical quartz vessel of 5 cm in diameter and 5 cm in height. The energy of the recoiled electron in the Compton scattering process was deposited in the LS, which is viewed by a PMT (XP2020) below. Seven coincidence detectors were placed 60 cm far away from the LS vessel. The coincidence detector consisted of an inorganic crystal scintillator (LaBr) and a PMT (XP2020). Signals from the PMTs were sent into a fan in-fan out (CAEN N625) and were then sent into the trigger board (CAEN N405) and the FADC (CAEN N6742).

\begin{center}
\includegraphics[width=7cm]{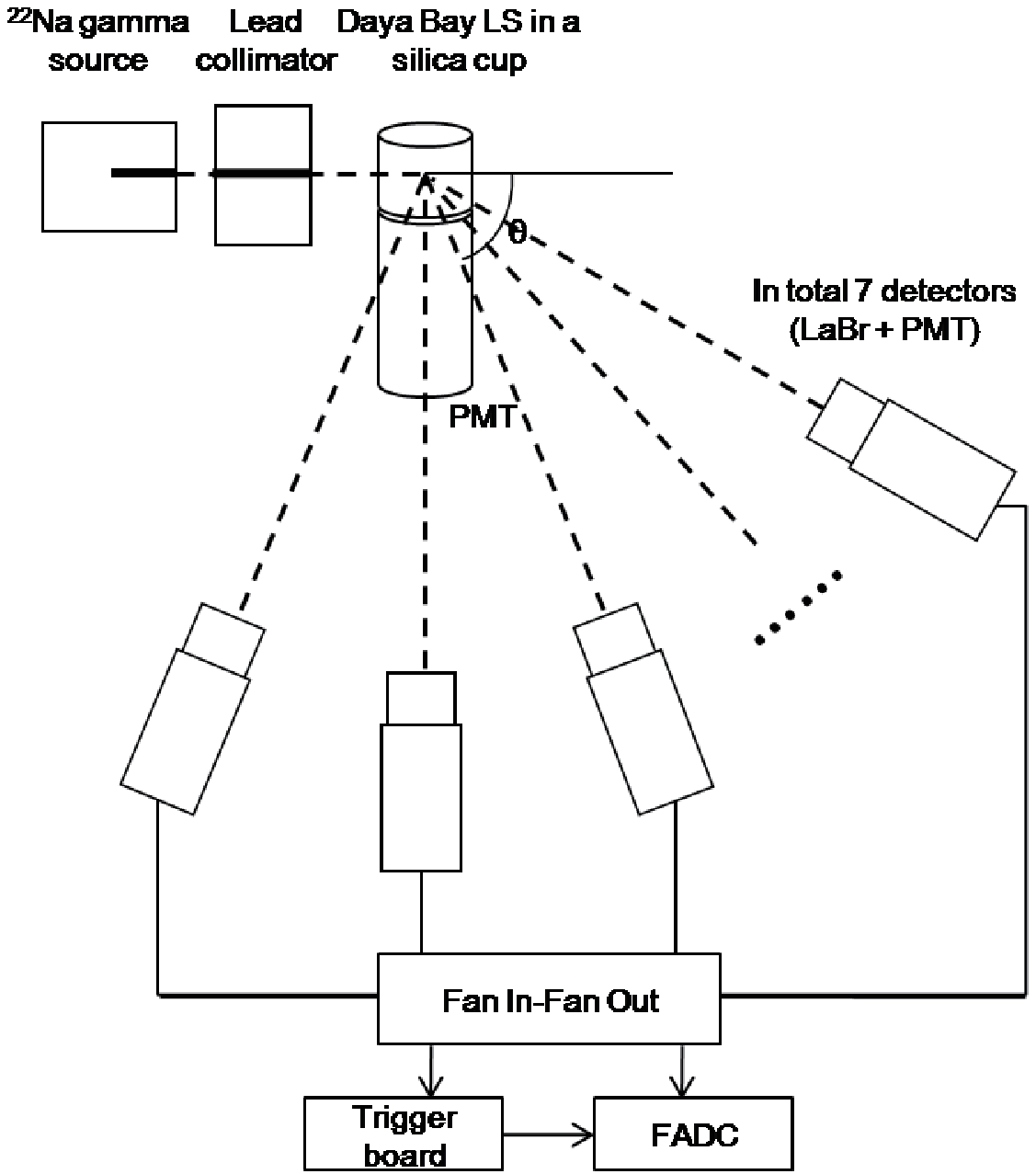}
\figcaption{\label{figsetup} Experimental setup. 7 detectors were placed around the LS to select Compton $\gamma$s. }
\end{center}

Seven coincidence detectors were put in seven directions (20\degree, 30\degree, 50\degree, 60\degree, 80\degree, 100\degree, 110\degree) and took data at the same time. The advantage of doing so is that we can avoid the influence of the possible system fluctuation when measuring at each directions one after another. The expected deposit energy of the recoil electron in the LS can be calculated with the Compton formula:
\begin{equation}\label{eqcompton}
  E_e = \dfrac{{E_{\gamma}}^2}{E_{\gamma}+\frac{m_e}{1-\cos \theta}}\,,
\end{equation}
where $E_e$ is the recoil electron energy, $E_\gamma$ is the $\gamma$ ray energy, $m_e$ is the electron mass, and $\theta$ is the Compton scattering angle.

\section{Data analysis}

\subsection{Pulse integration}

A typical FADC readout of the PMT under the LS vessel is showed in Fig.~\ref{figpulse}. The pulse was featured with steep rising edge and fast recovery. The baseline fluctuated within $\pm$5 channel. In this analysis, we used the average value of the first 50 readouts in each event as the baseline. The pulse threshold was defined as 20 FADC values lower than baseline. The pulse charge integrated from the readout passing the threshold to the readout back to baseline.

\begin{center}
\includegraphics[width=7cm]{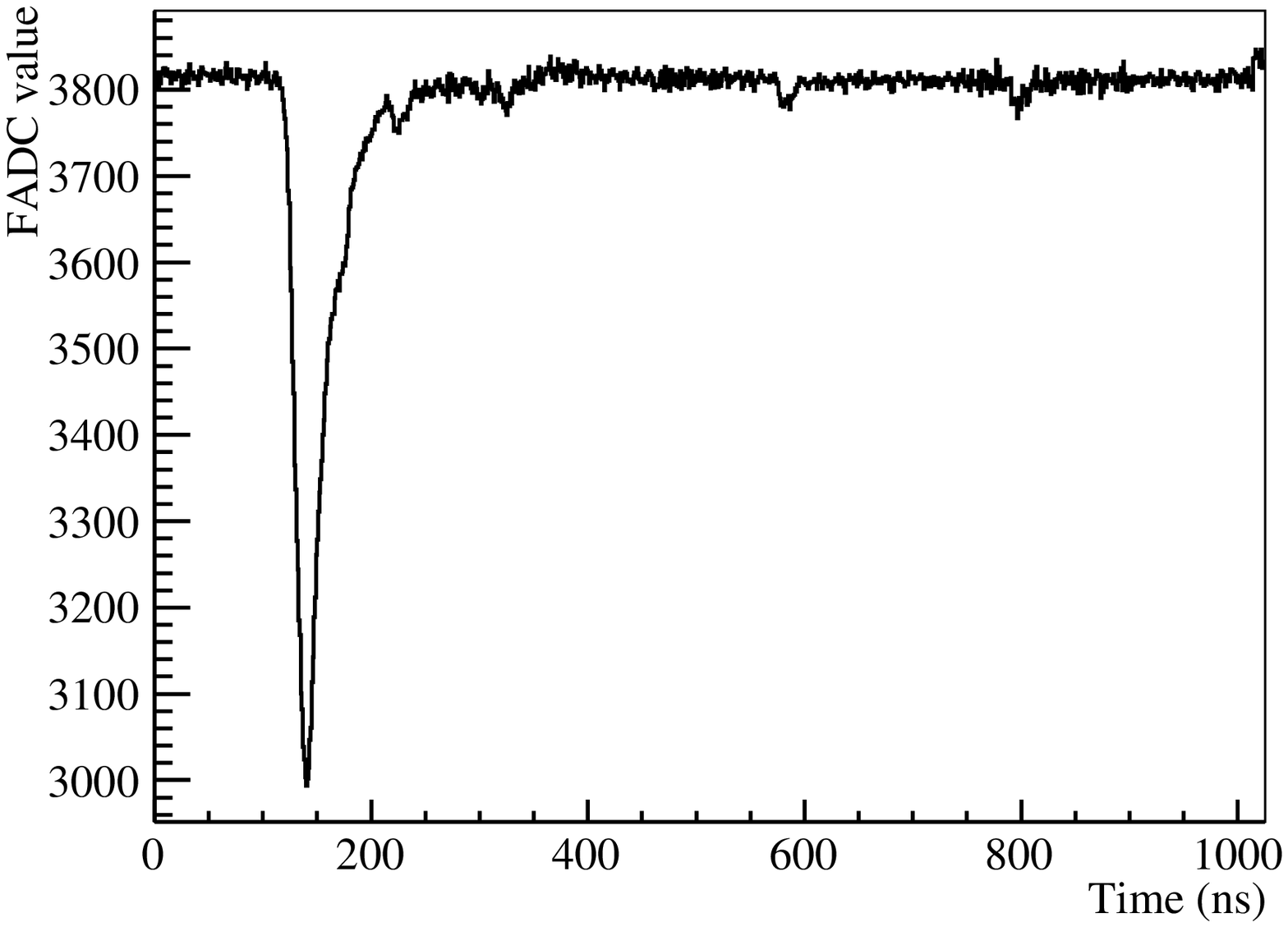}
\figcaption{\label{figpulse}An example of the PMT pulse shape recorded by the FADC.}
\end{center}

\subsection{Event selection}

The Compton scattering events were selected with the criteria that follows. First, events whose baseline had more than 5 FADC values deviation from the average baseline value during the whole measurement were rejected. Then if the pulse charge integration was smaller than 1000 FADC values, the event would most likely be a noise instead of a physics event, thus it was also rejected. A multiplicity cut was applied that if more than one coincidence detector had pulse passed threshold, the event would be rejected. The Compton electron and $\gamma$ pair was then selected by requiring the time interval between the two triggers $\Delta t = t_{e} - t_{\gamma}$ no more than 10 ns deviated from the average value. Fig.~\ref{figdt} showed the distribution of the time interval.

\begin{center}
\includegraphics[width=7cm]{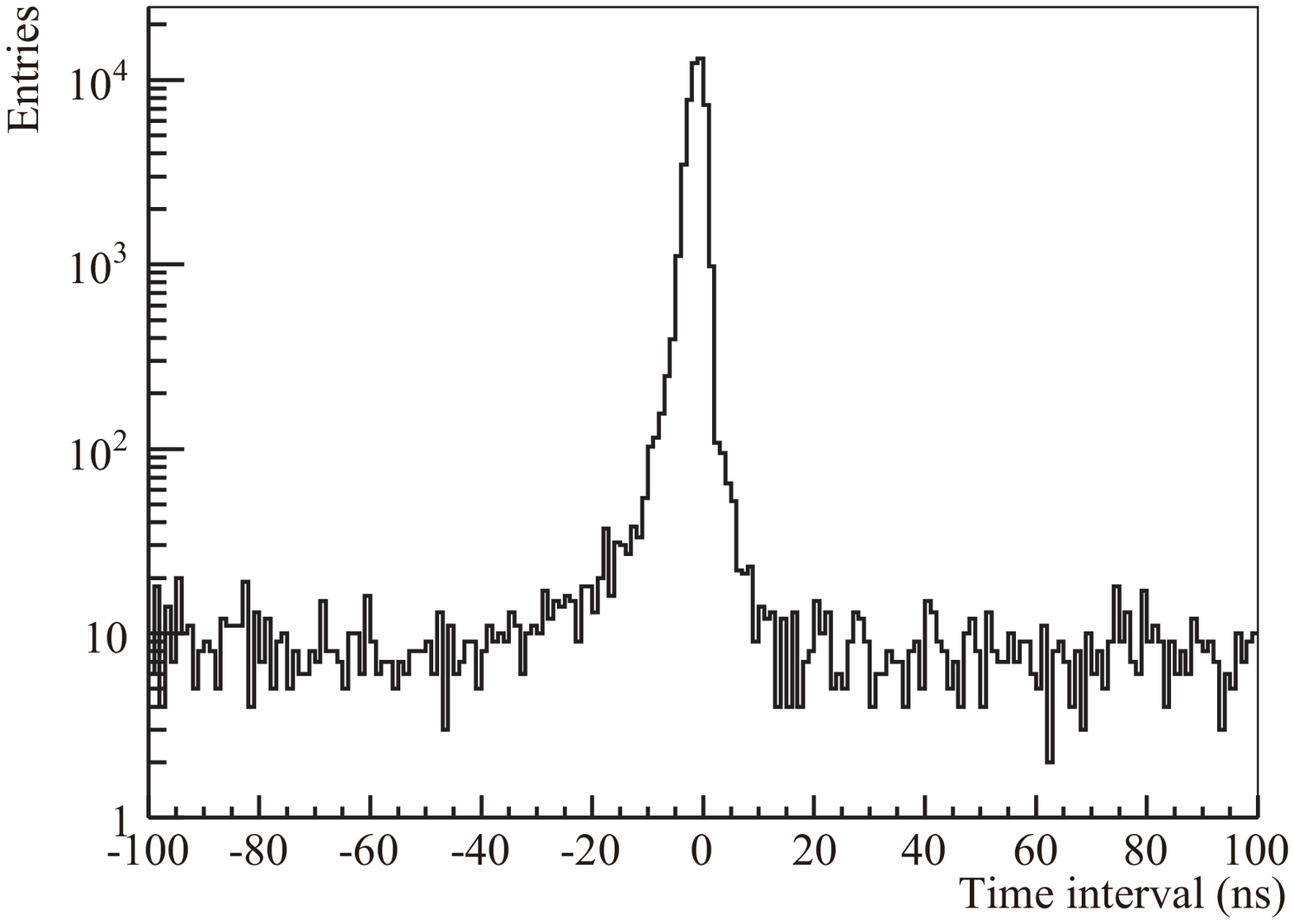}
\figcaption{\label{figdt} The distribution of the time interval between the coincidence triggers}
\end{center}

Fig.~\ref{figexample}(a) was an example of the energy correlation between events in LS and coincidence detector after the selection described above. Six zones were tagged on the figure. Zone A and B were Compton electron-$\gamma$ pairs of incident energy 0.511 MeV and 1.275 MeV respectively, which were needed for this analysis. Zone C and D were also Compton pairs except that the $\gamma$s scattered again in the coincidence detector. Zone E and F stood for multiple Compton scattering events in LS. With a group of coincidence detector energy selections, we could pick out events in Zone A and B. Examples of 1.275 MeV $\gamma$ spectra and 0.511 MeV $\gamma$ spectra were showed in Fig.~\ref{figexample}(c) and (d) respectively. Fitting results were also showed on these figures. We noticed the energy spectra were asymmetric, possibly due to energy leaks in the LS. So we used Crystal Ball function in fittings, which gave a good description for energy leaks \cite{gaiser}. Also Gaussian function was used as an alternative fitting function to estimate the systematical uncertainty of the fitting.

Accidental backgrounds were studied with the off-window time cut, requiring the time interval $\Delta t$ deviated from the average value within (10 ns, 100 ns). The background energy correlation between the LS and the coincidence detector was showed in Fig.~\ref{figexample}(b). The accidental backgrounds were mainly low energy events, and only very few were in the energy range of the coincidence events. The backgrounds in the samples were estimated to 0.01\%. It's influence on the energy peak was negligible.

\end{multicols}

\ruleup
\begin{center}
\includegraphics[width=12cm]{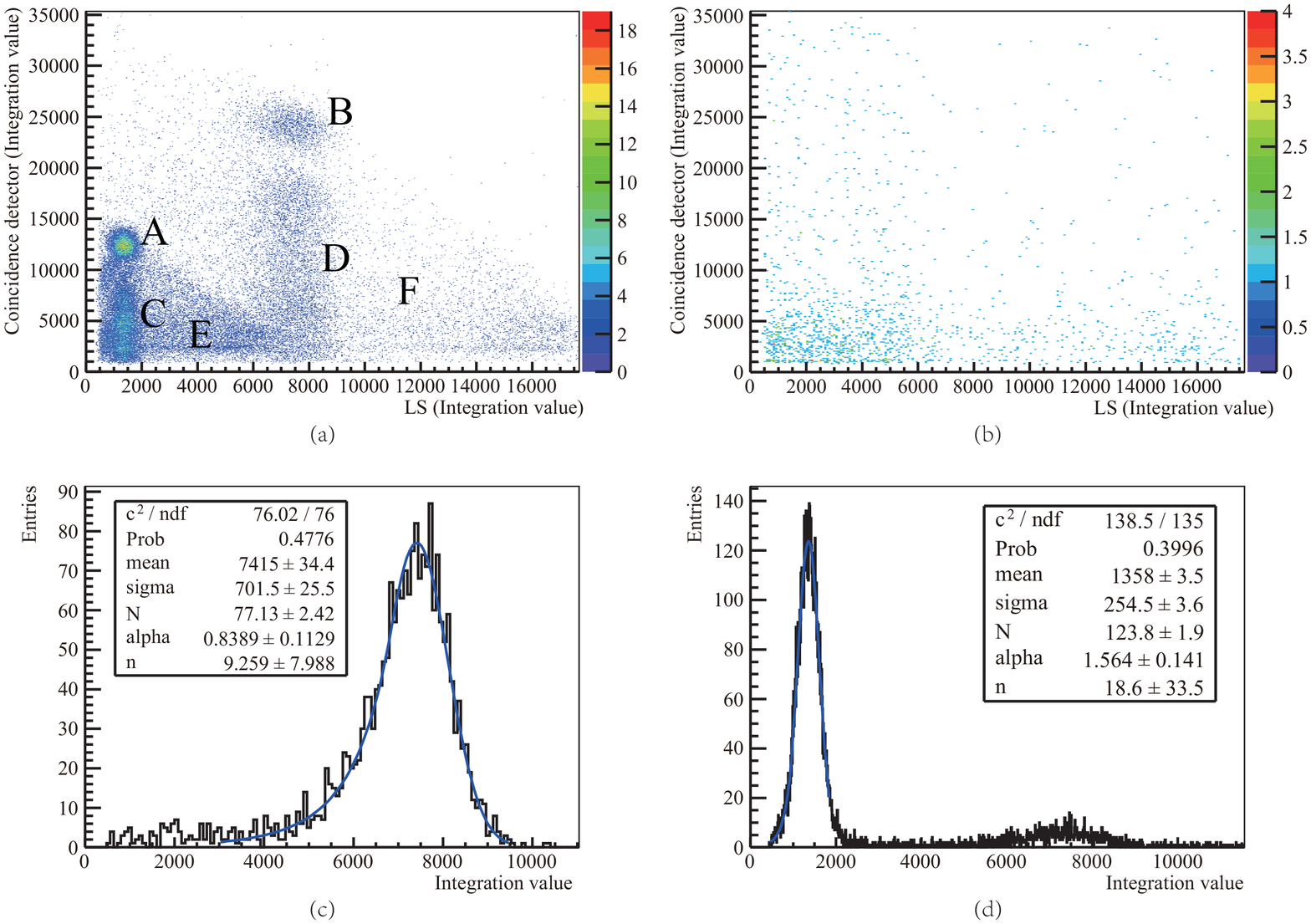}
\figcaption{\label{figexample} (a)Example of the energy correlation of coincident events. (b)Example of the energy correlation of accidental backgrounds selected by the off-window time cut. (c)Example of the 1.275 MeV $\gamma$ peak with fitting results. (d)Example of the 0.511 MeV $\gamma$ peak with fitting results}
\end{center}
\ruledown

\begin{multicols}{2}

\subsection{Systematical uncertainty estimation}

The major systematical uncertainty comes from the misalignment. Based on the geometrical survey, the maximum misalignment was estimated to 0.5\degree. Its influence on the energy response was angle dependent as listed in Table~\ref{tabrslt}, and was smaller than 1\% at most of the angles. In our measurement, the angle dispersion was $\pm$5\degree, which may also bias the scattering angle. The events with smaller scattering angles have larger possibility than those with larger scattering angles, which would induce systematical bias when the angle dispersion was too large. we developed a Monte-Carlo simulation to study the influence of the angle dispersion. It turned out that all angles were biased by 0.03\%$\pm$0.02\%, which was negligible in our case.

The nonlinearity induced by FADC was tested with the help of a pulse generator. The experimental setup was showed in Fig.~\ref{figfadcset}(a). We use a pulse generator to sent a rectangle pulse into the FADC. The energy response was then defined as $r_{FADC}=\frac{FADC\ integration\ value}{Pulse\ amplitude}$ To ensure no extra nonlinearity was induced by the pulse generator, an oscilloscope was also used to crosscheck the amplitude of the pulse. The results were showed in Fig.~\ref{figfadcset}(b). The nonlinearity induced by the FADC was conservatively estimated to 1\%.

\begin{center}
\includegraphics[width=7cm]{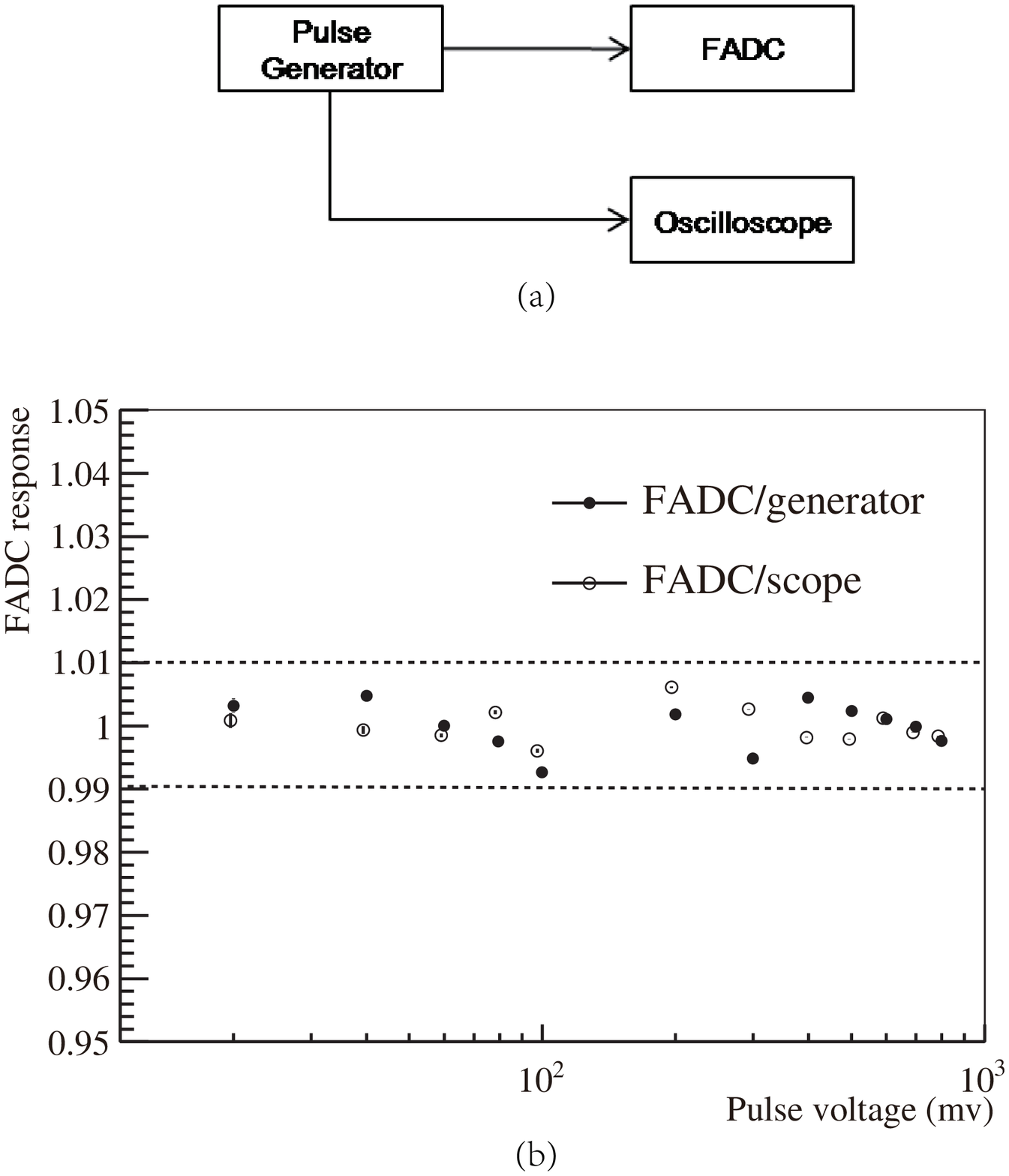}
\figcaption{\label{figfadcset} (a)FADC linearity measurement setup. An oscilloscope was also used to verify the linearity of the pulse generator. (b)FADC linearity measurement results. A band of $\pm 1\%$ was drawn to show the deviation.}
\end{center}

The systematical error of fitting was studied by comparing different fitting functions, and it was also 1\%. The PMT nonlinearity was a function of the anode peak current\cite{photonis}. We kept the PMTs working in the linear range by setting a relatively low working voltage. Its nonlinearity was estimated less than 1\%.

FADC nonlinearity, PMT nonlinearity and fitting uncertainty were partially correlated among the data points. However, for simplicity, we treat them as uncorrelated uncertainties.

\subsection{Electron energy response of the liquid scintillator}

The LS energy response to the true electron energy $E_{true}$ can be expressed as $R(E_{true})=E_{vis}/E_{true}$, where $E_{vis}$ was the visible light in LS. We used 1.275 MeV $\gamma$ peak at 80\degree as the normalization point, and its $R(E_{true}^{80, 1.275MeV})$ was anchored at 1. Then the energy response at $i$ degree of incident energy $E_j$ can be calculated by

\begin{equation}\label{eqlsresponse}
  R(E_{true}^{i, E_j}) = \dfrac{FADC^{i, E_j}/FADC^{80, 1.275MeV}}{E_{true}^{i, E_j}/E_{true}^{80, 1.275MeV}}
\end{equation}

in which $FADC^{i, E_j}$ was the fitted peak position in FADC value, and $E_{true}^{i, E_j}$ was calculated by Eq.~(\ref{eqcompton}).

The results were listed in Table~\ref{tabrslt}. Statistical uncertainties and misalignment uncertainties were also listed since they were angle dependent. FADC nonlinearity, PMT nonlinearity and fitting uncertainties were all 1\% for all angles as mentioned above.

\end{multicols}

\begin{center}
\tabcaption{\label{tabrslt}The electron energy response of LS and uncertainties. The FADC nonlinearity, PMT nonlinearity and fitting uncertainty were all 1\% for all angles, so they were not listed in this table.}
\footnotesize
\begin{tabular*}{170mm}{c@{\extracolsep{\fill}}ccccc}
\toprule Angle  &  True Energy (MeV)  &  $R(E_{true}^{i, E_j})$  &  Total Uncertainty  &  Stat. Uncertainty  &  Misalignment  \\
\hline
1.275 MeV $\gamma$ \\
20\degree & 0.167 & 0.823 & 4.7\% & 0.9\% & 4.3\% \\
30\degree & 0.319 & 0.947 & 3.1\% & 0.6\% & 2.4\% \\
50\degree & 0.601 & 0.975 & 2.1\% & 0.6\% & 1.0\% \\
60\degree & 0.707 & 0.996 & 1.9\% & 0.6\% & 0.7\% \\
80\degree & 0.858 & 1.000 & 1.9\% & 0.6\% & 0.3\% \\
100\degree & 0.950 & 0.991 & 1.8\% & 0.6\% & 0.2\% \\
110\degree & 0.982 & 0.970 & 1.8\% & 0.5\% & 0.1\% \\
0.511 MeV $\gamma$ \\
20\degree & 0.029 & 0.681 & 5.1\% & 1.1\% & 4.7\% \\
30\degree & 0.060 & 0.796 & 3.4\% & 0.6\% & 2.9\% \\
50\degree & 0.134 & 0.860 & 2.3\% & 0.5\% & 1.4\% \\
60\degree & 0.170 & 0.894 & 2.1\% & 0.5\% & 1.0\% \\
80\degree & 0.231 & 0.923 & 1.9\% & 0.5\% & 0.6\% \\
100\degree & 0.276 & 0.936 & 1.8\% & 0.5\% & 0.3\% \\
110\degree & 0.293 & 0.945 & 1.8\% & 0.5\% & 0.3\% \\
\bottomrule
\end{tabular*}
\end{center}

\begin{multicols}{2}

\section{Conclusion and discussion}

We had measured the electron response of the Daya Bay LS in a range of 0.03 MeV to 1 MeV through Compton scattering process, with uncertainties of 1.8\% (for large scattering angles) to 5\% (for small scattering angles). Tagging the scattered gamma from a collimated source significantly improves the precision of the energy measurement. Taking data with seven coincidence detectors simultaneously at different angles cancel out the instability of the system. Fig.~\ref{fignlresult} shows the measured energy response, with a fit using the empirical model used in Ref.~\cite{dyb_shape}.

In a large scale liquid scintillator detector, such as Daya Bay and JUNO, the energy nonlinearity is often a combination of the liquid scintillator and other effects, and is particle dependent. For example, in Daya Bay both liquid scintillator and readout electronics contribute. Determining the positron energy (and thus the derived neutrino energy) nonlinearity with in-situ calibration data of $\gamma$s and $\beta$-decays will carry relative large uncertainties due to the strong correlation between the liquid scintillator nonlinearity and the electronics nonlinearity. In Ref.~\cite{dyb_shape}, the positron nonlinearity was determined with an uncertainty $\sim 1.5\%$ for most of the relevant energy region. However, the liquid scintillator nonlinearity alone is $\sim10\%$ due to the correlation. The direct measurement of the liquid scintillator nonlinearity for electron in this note, combining with the in-situ calibrations, will significantly improve the precision of the nonlinearity of the Daya Bay detector.

\begin{center}
\includegraphics[width=7cm]{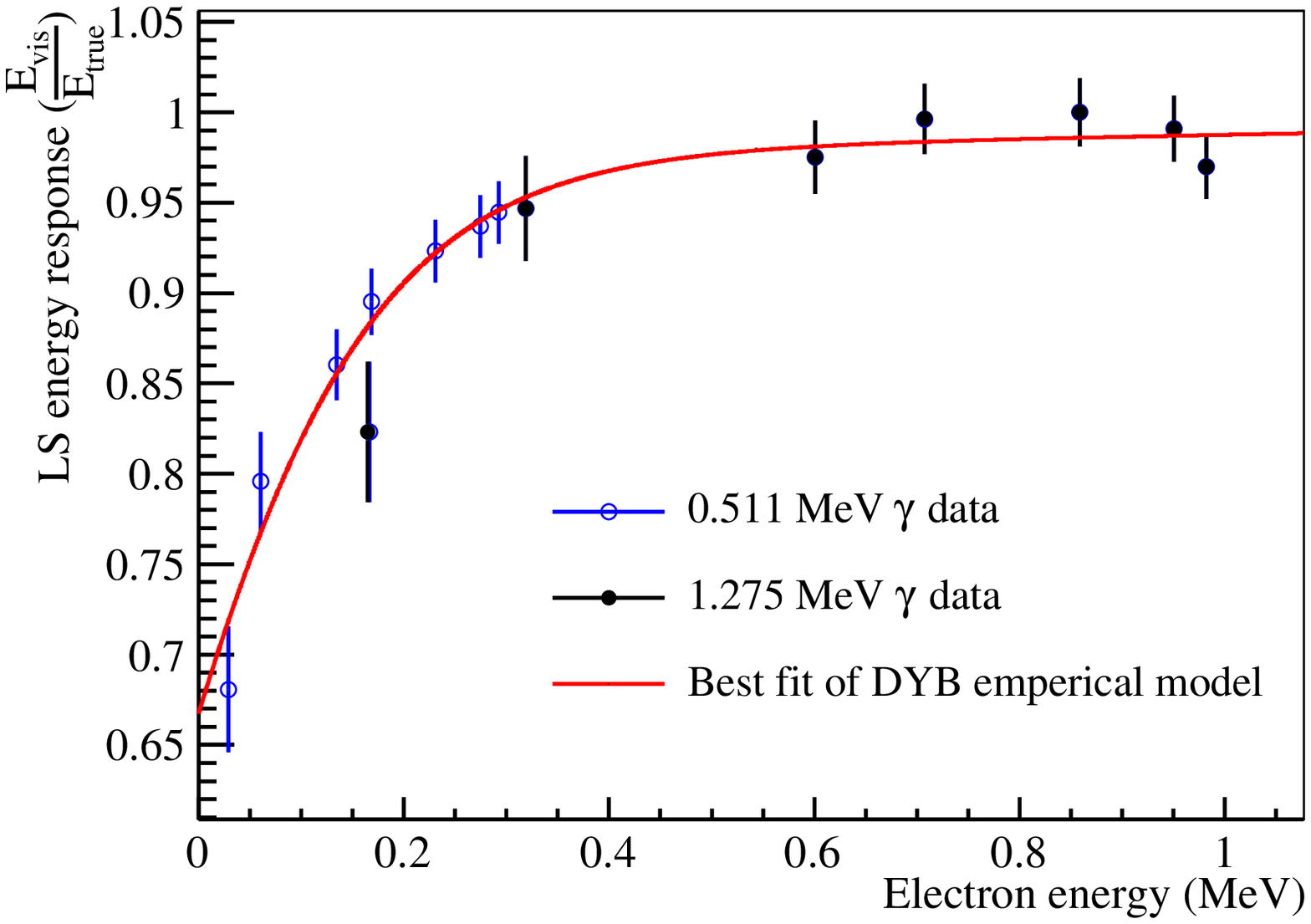}
\figcaption{\label{fignlresult} Electron energy response of the Daya Bay liquid scintillator. The solid line is the best fit of the empirical nonlinearity model.}
\end{center}

\end{multicols}

\vspace{-1mm}
\centerline{\rule{80mm}{0.1pt}}
\vspace{2mm}

\begin{multicols}{2}

\end{multicols}

\clearpage

%\end{CJK*}
\end{document}